\documentclass[letterpaper,english,aps,prl,twocolumn,floatfix, showpacs, amsfonts, amssymb]{revtex4}
\usepackage{epsfig, graphicx,psfrag,amsmath,amssymb,float}
\usepackage[T1]{fontenc}
\usepackage[latin9]{inputenc}
\usepackage{amsmath}
\usepackage{amssymb}
\usepackage{amscd}
\usepackage{bm}
\usepackage{psfrag}
\usepackage{babel}
\usepackage{graphicx}

\newcommand{\comment}[1]{}

\begin{document}
\title{Two-channel Kondo physics in two-impurity Kondo models}

\author{Andrew K. Mitchell,$^{1}$ Eran Sela,$^{1}$ and David E. Logan$^{2}$}

\affiliation{$^{1}$Institute for Theoretical Physics, University of
  Cologne, 50937 Cologne, Germany\\
$^2$Department of Chemistry, Physical and Theoretical
  Chemistry, Oxford University, South Parks Road, Oxford OX1 3QZ,
  United Kingdom}


\begin{abstract}
We consider the non-Fermi liquid quantum critical state of the spin-$S$ two-impurity Kondo model,
and its potential realization in a quantum dot device. Using conformal field theory (CFT) and the numerical renormalization group (NRG), we show the critical point to be identical to that of the two-channel Kondo model with additional potential scattering, for any spin-$S$. Distinct conductance signatures are shown to arise as a function of device asymmetry; with the `smoking gun' square-root behavior, commonly believed to arise at low-energies, 
dominant only in certain regimes.
\end{abstract}

\maketitle


Systems comprising several quantum impurities inherently display an interplay
between impurity-bath and interimpurity couplings~\cite{2ik:RKKY,2ik:jones_prl1,2ik:aff_lud,2ik:aff_lud_jones,gan,zarandPRL,Chang_rev}:
the tendency for impurities to be Kondo-screened by conduction electrons competes with 
screening by interimpurity spin-singlet formation. Rich physics can thereby arise, 
as demonstrated for example in coupled quantum
dots~\cite{dd_rev,Chang_rev}, magnetic impurities in metals~\cite{2ik:RKKY}, 
and recent two-impurity STM experiments~\cite{Bork2}. Indeed, the same essential
physics governs the analogous propensities for heavy Fermion behavior
or magnetic ordering in lattice systems~\cite{hewson}.

The two-impurity, spin-$\tfrac{1}{2}$ Kondo model (2IKM) is the simplest to capture this 
competition~\cite{2ik:RKKY}: local singlet formation is favored by an 
interimpurity exchange $K$, while coupling of each impurity to its own conduction
channel favors separate Kondo screening below an effective single-channel, single-impurity scale $T_K$~\cite{2ik:jones_prl1,2ik:aff_lud,2ik:aff_lud_jones,gan}. 
The lack of interchannel charge transfer in the 2IKM permits two
distinct phases, and a quantum phase transition (QPT) results on
tuning $K_c\sim T_K$. At the critical point, non-Fermi liquid (NFL)
physics arises below $T_c$, characterized~\cite{2ik:aff_lud} by
anomalous properties such as fractional residual entropy and singular
magnetic susceptibility.

This critical physics is also surprisingly robust to some perturbations,
notably breaking of mirror (parity) symmetry or particle-hole
symmetry~\cite{zarandPRL}. Such perturbations are marginally
irrelevant in the sense that the interimpurity coupling can be retuned to recover the
critical point in all cases. But despite considerable effort, 2IKM critical physics has
proved experimentally elusive --- mainly due to interchannel charge transfer
which smooths the QPT into a crossover~\cite{2ik:aff_lud_jones}. Regular Fermi liquid (FL) 
physics then sets in below an energy scale $T^*$; and if 
the degree of charge transfer is large enough that $T^*\gg T_c$, no 
evidence of the critical point will be observed~\cite{DEL2iam}. This
is the situation relevant to the recent two-impurity experiments of
Ref.~\cite{Bork2}: coupling between one impurity on a metal surface and
one on an STM tip was also accompanied by strong tip-surface tunneling.

Reducing the degree of interchannel charge transfer might be possible
in a quantum dot device such as that proposed in Ref.~\cite{zarandPRL}. 
Provided $T^*\ll T_c$, NFL behavior should be observable in an
intermediate energy window, as can be understood from a 2IKM critical
perspective (indeed the eventual crossover to FL physics is wholly
characteristic of the intermediate NFL state~\cite{akm_2ck2ik}).

An alternative route could however involve use of a quantum box,
which acts as an interacting lead~\cite{2ck_expt}. Coupling a 
single dot to one regular lead and one box tuned to the Coulomb 
blockade regime, suppresses interchannel charge transfer completely.
This has been exploited to access single-impurity two-\emph{channel} Kondo (2CK) physics~\cite{2ck:nozieres,2ck:rev_cox_zaw} in a real device~\cite{2ck_expt}.
Here we propose simply to interject a second dot in series between the `leads' to 
realize 2IKM physics. While parity and particle-hole symmetries are thereby broken, 
the QPT itself is unaffected. Robust NFL behavior should persist down to the lowest 
energy scales at the critical point.

Here we address two key questions in regard to potential realization of 2IKM physics. First, 
what is the nature of the critical point itself? We show that it is \emph{identical} to that 
arising in a 2CK model with additional potential scattering, \emph{independent} of parity breaking. 
Further, we show that the same QPT and 2CK critical point arises in the spin-$S$ generalization of the 
2IKM. Second, what are the signatures of criticality in measurable quantities such as conductance?
These reflect RG flow from higher-energy fixed points (FPs), and depend sensitively on parity
breaking. We find in particular that the square-root behavior commonly anticipated~\cite{zarandPRL,Chang_rev}
at low-energies, is absent in the standard channel-symmetric 2IKM.


\emph{Nature of 2IKM critical point.--} The 2IKM reads:
\begin{equation}
\label{H2ik}
H_{2IK} =H_0+H_{\textit{ps}}+J_{L}  \vec{S}_L \cdot \vec{s}_{0L} + J_{R}\vec{S}_R
\cdot \vec{s}_{0R} +K \vec{S}_L \cdot \vec{S}_R ,
\end{equation}
where $H_{0}=\sum_{\alpha,k} \epsilon_{k}^{\phantom{\dagger}} 
\psi_{k\sigma \alpha}^{\dagger} \psi^{\phantom{\dagger}}_{k \sigma\alpha}$ 
describes two free conduction channels $\alpha=L/R$, with density of states $\rho$,
and spin density at the impurities $\vec{s}_{0\alpha}=\sum_{\sigma\sigma'}\psi_{0\sigma \alpha}^{\dagger}
(\tfrac{1}{2}\vec{\sigma}_{\sigma\sigma'})\psi^{\phantom{\dagger}}_{0
 \sigma' \alpha}$  (where $\psi_{0\sigma \alpha}^{\dagger}=
\sum_{k}\psi_{k\sigma \alpha}^{\dagger}$). Potential scattering is included via 
$H_{\textit{ps}}=\sum_{\alpha} V_{\alpha}\psi_{0 \sigma \alpha}^{\dagger} \psi_{0 \sigma
 \alpha}^{\phantom{\dagger}}$, and   
$\vec{S}_{\alpha}$ are spin-$\tfrac{1}{2}$ 
operators for the impurities. The 2IKM has been extensively studied using a number of powerful
techniques~\cite{2ik:jones_prl1,2ik:aff_lud,2ik:aff_lud_jones,gan,akm_2ck2ik}, 
and certain similarities have been found~\cite{2ik:aff_lud_jones,zarandPRL,gan,akm_2ck2ik,DEL2iam} 
between it and the 2CK model~\cite{2ck:nozieres,2ck:rev_cox_zaw}, 
\begin{equation}
\label{H2ck}
H_{2CK} =H_0+H_{\textit{ps}}+J_{L}  \vec{S} \cdot \vec{s}_{0L} + J_{R}\vec{S}
\cdot \vec{s}_{0R} ,
\end{equation}
with $\vec{S}$ a spin-$\tfrac{1}{2}$ operator for a single
impurity, exchange-coupled to two independent conduction channels. The
physics of the 2CK model is itself immensely 
rich~\cite{2ck:rev_cox_zaw}: the impurity is fully Kondo
screened by the more strongly-coupled conduction channel below
an effective single-channel scale $T_K$, producing two distinct phases
as a function of $(J_L-J_R)$. But when $J_L=J_R$, the
frustration inherent when two channels compete to screen the impurity
results in `overscreening', and NFL physics results below
$T_K^{2CK}$~\cite{2ck:rev_cox_zaw}. Strikingly, both 2CK and 2IKM have 
the same fractional residual entropy, 
$S_{\text{imp}}=\tfrac{1}{2}\ln(2)$.

\begin{figure}[t]
\begin{center}
\includegraphics*[width=80mm]{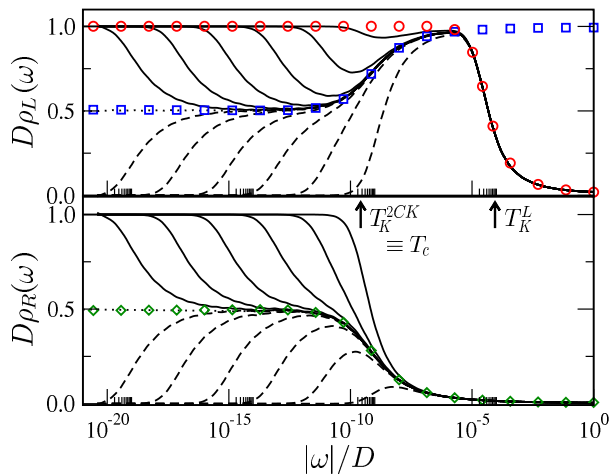}
\caption{\label{fig:spec}
Spectra $D\rho_{\alpha}(\omega)$ vs frequency
$|\omega|/D$ for channels $\alpha=L$ and $R$ [upper and lower panels] for the 
asymmetric 2IKM with fixed Kondo exchanges $\rho J_L=0.1$, $\rho J_R=0.05$,
varying interimpurity exchange $K$ (and $H_{\textit{ps}}=0$). Plotted using 
$K=K_c(1\pm 10^{-n})$ for dashed and solid lines, with integer $n=1\rightarrow 6$ 
approaching progressively the critical point $K_c\sim T_{K}^{L}\approx 10^{-6}D$ 
(dotted line). Circles: for a 1CK model with $\rho J=\rho J_L$. Diamonds: 
$D\rho_{2CK}(\omega)$ for a pure 2CK model with $T_K^{2CK}=T_c$.
Squares: $D\tilde{\rho}_{2CK}(\omega)=1-D\rho_{2CK}(\omega)$.
}
\end{center}
\end{figure}

CFT has been used to describe the critical points of both models~\cite{2ik:aff_lud_jones,2ck:aff_lud}.
In the `unfolded' representation, $H_0$ is written in terms of left-moving chiral Dirac Fermions. 
The CFTs for each model can be separated into 
different symmetry sectors. 
In particular, the $SU(2)_2 \times SU(2)_2
\times U(1)$ flavor, spin and charge symmetries of the 2CK model~\cite{2ck:aff_lud}, and the 
$U(1) \times U(1) \times SU(2)_2 \times Z_2$ left/right charge, total spin and Ising symmetries of
the 2IKM~\cite{2ik:aff_lud} can be exploited. The 2CK and 2IKM critical fixed point
Hamiltonians take the same form as $H_0$, but with modified boundary conditions (BCs) that affect 
only the spin sector of the 2CK model~\cite{2ck:aff_lud} or the Ising sector of the
2IKM~\cite{2ik:aff_lud_jones}. The finite size spectrum (FSS) at the critical point of each model,
can then be determined~\cite{2ik:aff_lud_jones,2ck:aff_lud}. In the channel-symmetric case $J_L=J_R$ 
and for $H_{ps}=0$, the FSS of the 2CK critical point is characterized by the fractions $0$,
$\tfrac{1}{8}$, $\tfrac{1}{2}$, $\tfrac{5}{8}$, $1$...; while for 2IKM a different FSS 
arises: $0$, $\tfrac{3}{8}$, $\tfrac{1}{2}$, $\tfrac{7}{8}$, $1$...

Despite these apparent differences between the critical FPs of the two models, the 2IKM can be 
mapped onto an effective 2CK model in special cases~\cite{gan,DEL2iam,zarandPRL}. The key 
requirement for that mapping is of course the generation of an effective spin-$\tfrac{1}{2}$ local 
moment (LM), which can then be overscreened by symmetric coupling to two conduction channels. 
In the channel-asymmetric limit $J_L\gg J_R$, 2CK critical physics arises via a simple 
mechanism~\cite{zarandPRL}, first involving Kondo screening of the $L$-impurity by the $L$-lead 
on the single-channel scale $T_K^L$; followed by second-stage overscreening of the $R$-impurity by 
the $R$-lead and an effective coupling to the remaining Fermi liquid bath states of the $L$-lead below $T_c$. 
An effective 2CK model of form Eq.~\ref{H2ck}, valid at low-energies $T\lesssim
T_K^{L}$, can be derived formally using the approach of Ref.~\cite{akm_oddchains}, exploiting the 
Wilson chain representation~\cite{NRG_rev}; and effective couplings follow as $\rho J^{\textit{eff}}_L\sim K/T_K^L$ and $\rho J^{\textit{eff}}_R\sim [1/\rho J_R - 1/\rho J_L]^{-1}$. The 2CK FP is thus stable when 
$K=K_c\sim T_K^L \rho J_R^{\textit{eff}}$, so the low-energy physics of the 2IKM is in this case wholly 
equivalent to that of the 2CK model.

Importantly however, the $L$-channel free electrons in the effective 2CK model acquire a $\pi/2$ phase 
shift due to the first-stage single-channel Kondo screening of the $L$-impurity in the original 2IKM. 
This is seen clearly in the dynamics of the asymmetric 2IKM; to demonstrate which, and to highlight the 
basic physical picture, Fig.~\ref{fig:spec} shows spectra 
$D\rho_{\alpha}(\omega)\equiv -\pi\rho\text{Im}[t_{\alpha}(\omega)]$
vs $|\omega|/D$ with $t_{\alpha}(\omega)$ the scattering t matrix~\cite{akm_frust}. 
Results are obtained from NRG, exploiting all model symmetries, discretizing conduction bands of width $2D$ logarithmically using $\Lambda=3$, and retaining $8000$ states per iteration in each of $z=3$ interleaved
calculations (for a review, see Ref. \cite{NRG_rev}).

Single-channel Kondo screening of the $L$-impurity by the $L$ lead on the scale of $T_K^L$ is seen 
directly in the $L$ spectra in the upper panel: a Kondo resonance, reaching the unitarity limit
$D\rho_{L}=1$, which has precisely the form of a regular single-channel Kondo (1CK) model~\cite{hewson}.
This embodies the $\pi/2$ phase shift in the $L$ channel; but no such feature is
observed on this energy scale in the $R$ spectra (lower panel),
indicating that the $R$-impurity is still essentially free. On tuning
the interimpurity coupling $K$ closer to the critical point of the 2IKM, 
the spectra in both channels fold progressively onto the
critical spectra. For energies $|\omega|/D \ll T_K^L$, the critical spectrum $D\rho_R(\omega)$ 
is precisely that of a 2CK model $D\rho_{2CK}(\omega)$ with $T_K^{2CK}=T_c$.
But to leading order the critical spectrum in the left channel is
$D\rho_{L}(\omega)=1-D\rho_{2CK}(\omega)$~\cite{akm_oddchains}. Thus, at the channel-asymmetric critical point~\cite{zarandPRL},
\begin{equation}
\label{tmasymptotic}
D\rho_{\alpha}(\omega) \overset{|\omega|\ll T_c}{\sim} \tfrac{1}{2}+\alpha\beta\sqrt{|\omega|/T_c}
\end{equation}
with $\alpha=\pm 1$ for channel $L/R$ resulting from the additional $L$-channel
$\pi/2$ phase shift, and $\beta$ a constant $\mathcal{O}(1)$.

This phase shift can be included in the 2CK model, Eq.~\ref{H2ck}, via the potential
scattering term $H_{\textit{ps}}$ (ie $V_L\rightarrow \infty$ but $V_R=0$, accompanied also by 
retuning  $J_L$ and $J_R$ to access the critical point). This is equivalent to adding 
infinite uniform and staggered potential scatterings, which affect respectively the
charge and flavor sectors of the 2CK model. Modifying the CFT for the
critical point of the 2CK model to include this, we find~\cite{sm} the BC becomes 
equivalent to that of the 2IKM. The FSS is also naturally affected and is given by~\cite{sm}
\begin{equation}
\label{cft_2ck_E}
E_{2CK}=\tfrac{1}{8}(Q-a)^2 + \tfrac{1}{4}j(j+1) +
\tfrac{1}{4}j_F(j_F+1) - bj_F^z,
\end{equation}
where $Q$, $j$ and $j_F$ are the charge, spin and flavor quantum
numbers. Uniform potential scattering shifts the charge parabolas,
while staggered potential scattering biases the 
flavor sector. The $\pi/2$ phase shift in the $L$-channel corresponds
to $a=1$ and $b=\tfrac{1}{2}$~\cite{sm}. Only certain quantum number
combinations are allowed at the critical point, as given by the
nontrivial gluing conditions derived in Ref.~\cite{2ck:aff_lud}; 
and which reproduce fully the 2IKM spectrum when used with
Eq.~\ref{cft_2ck_E} (see Fig.~2 of~\cite{sm}). One remarkable result
obtained from our NRG calculations~\cite{sm} is that the FSS at the
critical point of the 2IKM does not depend on channel asymmetry  
(whence in particular the critical point possesses an emergent parity
symmetry, irrespective of bare model symmetries). Further,
we have shown~\cite{sm} that the critical point for one 
model with potential scattering $V_L$ and $V_R$ is equivalent to the
critical point of the other model with different potential scattering
$\tilde{V}_L$ and $\tilde{V}_R$. 
The 2IKM and 2CK critical FPs are thus equivalent in the sense that they lie on the same
marginal NFL manifold parametrized by $H_{ps}$.


\begin{figure}[t]
\begin{center}
\includegraphics*[width=70mm]{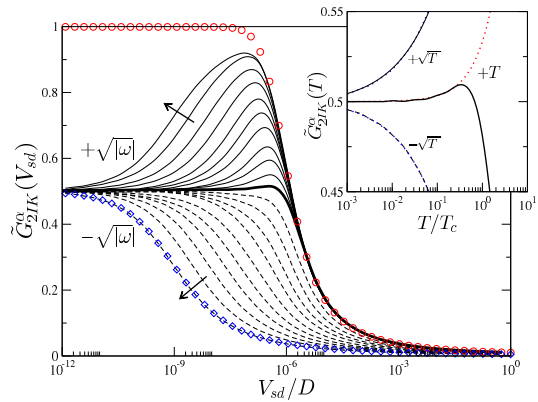}
\caption{\label{fig:cond}
$T=0$ conductance $G_{2IK}^{\alpha}/(2e^2h^{-1}G_0^{\alpha})$ through channel $\alpha=L$
and $R$ (solid and dashed lines) vs bias $V_{sd}/D$, at the critical point.
Shown for $\rho J_L = 0.075 \ge \rho J_R$, varying $\rho J_R=0.075 \rightarrow
0.05$ in steps of $0.0025$, with $K=K_c \sim T_{KL}^{1CK}$ retuned in
each case (and $H_{\textit{ps}}=0$).  Thick solid line is the symmetric case $J_L=J_R$; 
asymmetry $J_L/J_R\ge 1$ increases in direction of arrows. 
Circle and diamonds: pure 1CK and 2CK scaling spectra.
Inset: zero-bias conductance vs $T/T_c$ for $J_L/J_R=1$ and $2$ (and $\alpha =L,R$), 
exhibiting respectively leading \emph{linear} and square-root behavior (dotted lines).
}
\end{center}
\end{figure}

\emph{Conductance lineshapes and symmetry.--} 
Full RG flow from the local moment (LM) FP to the 2CK FP is thus recovered at the critical point
of the asymmetric 2IKM. This is manifest~\cite{zarandPRL} in the conductance arising 
e.g.\ when a given channel $\alpha=L/R$ is split into source and drain. At zero-bias, it is given
exactly~\cite{mw} in terms of the scattering t matrix (considered for
the channel-asymmetric case in Fig.~\ref{fig:spec}) by
$G_{2IK}^{\alpha}(V_{sd}=0,T)/(2e^2h^{-1}G_0^{\alpha})=-\int_{-\infty}^{\infty}
d\omega \partial f(\omega/T)/\partial \omega D\rho_{\alpha}(\omega,T)$;
with $f(\omega/T)$ the Fermi function, and the impurity-lead coupling parametrized by
$G_0^{\alpha}=4\Gamma_s^{\alpha}\Gamma_d^{\alpha}/(\Gamma_s^{\alpha}+\Gamma_d^{\alpha})^2$,
in terms of the $\alpha=L/R$ hybridizations to source ($\Gamma_s^{\alpha}$) and drain ($\Gamma_d^{\alpha}$).
Indeed, in the limit $\Gamma_s^{\alpha}\gg \Gamma_d^{\alpha}$ (ie. $G_0^{\alpha}\ll 1$), the $T=0$ conductance
follows as $\tilde{G}_{2IK}^{\alpha}(V_{sd}) =G_{2IK}^{\alpha}(V_{sd},T=0)/(2e^2h^{-1}G_0^{\alpha})=D\rho_{\alpha}(\omega=V_{sd},T=0)$, 
and hence from Eq.~\ref{tmasymptotic} one finds at low energies $V_{sd}\ll T_c$,
\begin{equation}
\label{cond}
\tilde{G}_{2IK}^{\alpha}(V_{sd})= \tfrac{1}{2}+\alpha\beta\sqrt{V_{sd}/T_c}+\gamma_{\alpha}^{\phantom\dagger}(V_{sd}/T_c)+...
\end{equation}
where we include also a term linear in $V_{sd}/T_c$.

The leading square-root behavior of Eq.~\ref{cond} has been viewed as the `smoking gun' 
signature of this 2CK physics~\cite{zarandPRL,2ck:cond}, and was used  
to identify the critical point in the 2CK experiment of Ref.~\cite{2ck_expt}.
But we note that, unlike the 2CK model, the 2IKM does not possess
$SU(2)$ flavor symmetry. Since symmetry dictates which operators
can act in the vicinity of the critical FP, this is naturally
reflected in the asymptotic  
conductance through the coefficients $\beta$ and
$\gamma_{\alpha}$. Indeed, the full  
energy-dependence of conductance depends on the unstable FPs, whose
vying effects on  
RG flow again depend on symmetry and model parameters.
For example, in the usual symmetric 2IKM (Eq.~\ref{H2ik} with $J_L=J_R$
and $H_{\textit{ps}}=0$), no incipient LM is formed: there is no
intermediate energy window with eg. $S_{\text{imp}}=\ln(2)$ entropy, and RG
flow proceeds directly to the 2CK FP from the LM$\times$LM high energy
FP describing a pair of free impurities ($S_{\text{imp}}=\ln(4)$).

The effect of parity-breaking is explored in Fig.~\ref{fig:cond}, showing 
NRG results for conductance vs bias $V_{sd}$ at the 2IKM critical
point (obtained 
for $G_0^{\alpha}\ll 1$, as above). Conductance in the asymmetric limit $J_{L} \gg J_{R}$ 
is consistent with Ref.~\cite{zarandPRL}, and physical expectation as above
(see Eq.~\ref{cond}). Here, the asymptotic conductance is 
$\tilde{G}_{2IK}^R(V_{sd})\simeq 1-\tilde{G}_{2IK}^L(V_{sd})\equiv \tilde{G}_{2CK}(V_{sd})$ at 
low energies $V_{sd}\ll T_K^L$, where $T_K^L\gg T_c\equiv T_K^{2CK}$ 
and with  $\tilde{G}_{2CK}(V_{sd})$ the conductance of the standard 2CK
model~\cite{2ck:cond,2ck_expt} (diamonds). Thus, on exchanging $J_R\leftrightarrow J_L$, 
the coefficient $\beta$ of Eq.~\ref{cond} must change sign. But what happens as the asymmetry 
is decreased? We find the leading square-root contribution in Eq.~\ref{cond} vanishes (see 
Fig.~\ref{fig:cond}), as $\beta \sim (J_L-J_R)$, and leading \emph{linear} behavior 
emerges at the symmetric point $J_L=J_R$ (the same naturally arising as a function of $T$ 
at zero-bias, see Fig.~\ref{fig:cond} inset). 
In fact linear-$V_{sd}$ behavior also emerges as the symmetric point is
approached, since the square-root term dominates over a shrinking window
$V_{sd}/T_c \ll (\beta/\gamma_{\alpha})^{2}$.

Another striking feature of the conductance in more channel-symmetric 
situations is the behavior at higher $T$ or energies $\gtrsim T_c
\approx T_K^L$. Here the behavior is wholly characteristic of
single-impurity, single-channel Kondo physics, as seen by comparison
to the circles in Fig.~\ref{fig:cond}.   

The absence of square-root behavior in conductance of the symmetric 2IKM 
is contrary to common belief~\cite{zarandPRL,Chang_rev},
so we sketch now our CFT proof~\cite{sm}.  
As pointed out in Refs.~\cite{2ik:aff_lud_jones,zarandPRL},  
corrections to the t matrix in the vicinity of the critical point
(whose $\omega$-dependence displays the same scaling as conductance)
are determined from irrelevant boundary operators consistent  
with symmetry. Two such play a role here: 
$\delta H_1=c_1\epsilon^{\prime}$ and $\delta H_2=c_2\vec{J}_{-1}\cdot\vec{\phi}$ (in the notation of Ref.~\cite{2ik:aff_lud_jones}).  The operator $\vec{J}_{-1}\cdot\vec{\phi}$ is the leading irrelevant operator of the 2CK FP, whose effect on the t matrix is known~\cite{CFT2CK} to yield the famous square-root behavior. However,  
$\vec{J}_{-1}\cdot\vec{\phi}$ has \emph{odd} parity in the 2IKM (unlike 2CK), which implies that its coefficient $c_2\sim (J_L-J_R)$ vanishes in the symmetric limit. In both models, $\delta H_1$ does still contribute ($c_1$ always being finite~\cite{2ik:aff_lud_jones}). One might naively expect $\epsilon^{\prime}$ to behave similarly to $\vec{J}_{-1}\cdot\vec{\phi}$ since they have the same scaling dimension $3/2$. However, the key difference between $\epsilon^{\prime}$  and $\vec{J}_{-1}\cdot\vec{\phi}$ 
is that only the latter is a Virasoro primary field. In consequence~\cite{sm}, the leading square-root correction to the t matrix from $\delta H_1$ vanishes. In the symmetric 2IKM, the leading square-root behavior of conductance thus also vanishes.


\begin{figure}[t]
\begin{center}
\includegraphics[width=80mm]{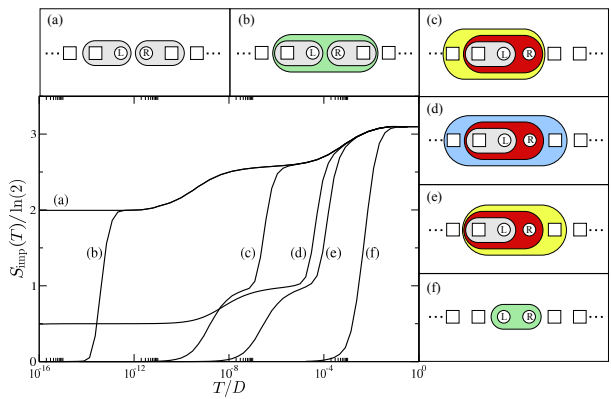}
\caption{\label{fig:s1}
Impurity entropy $S_{\text{imp}}(T)$ vs $T/D$ for the 2IKM with
spin-1 impurities. Plotted for fixed $\rho J_L=0.15$, $\rho J_R =
0.05$, varying $K/D=0$, $10^{-13}$, $3\times 10^{-7}$,
$K_c\approx 6\times 10^{-5}$, $2\times 10^{-4}$ and $10^{-2}$ for lines
(a)--(f). The corresponding physical processes are illustrated in
panels (a)--(f): impurities denoted as circles and conduction band
orbitals (in the Wilson chain representation~\cite{NRG_rev}) as
squares. For discussion, see text.
}
\end{center}
\end{figure}

\emph{Spin-S 2IKM.--}
Multilevel quantum dots can behave like $S=1$ impurities~\cite{DELmultilevel}, and 
high-spin impurities such as Co ($S=3/2$) have been manipulated with STM~\cite{co_stm}. 
Thus a natural and pertinent generalization of the 2IKM involves spin-$S$ impurities: the 
model remains Eq.~\ref{H2ik}, but $\vec{S}_{R},\vec{S}_{L}$ are now spin-$S$
operators.

A QPT must again arise, as follows from the same line of argument as the 
spin-$\tfrac{1}{2}$ 2IKM~\cite{2ik:aff_lud}. On tuning $K$ there is a
phase-shift 
discontinuity on going from the local singlet phase for large $K$ to a 
separated spin-$S$ \emph{underscreened} Kondo phase for small $K$. 
The nature of the transition arising at $K_c$ is again clear by considering
the asymmetric limit $J_L\gg J_R$. For concreteness consider $S=1$,
although the argument  
extends easily to higher-$S$. NRG results for the entropy
$S_{\text{imp}}(T)$ vs $T$ are shown in Fig.~\ref{fig:s1}, together
with cartoons highlighting the key physical processes. In (a)
the impurities are completely decoupled ($K=0$), with each thus
underscreened to a spin-$\tfrac{1}{2}$ by its attached lead $\alpha$, on its own
single-channel Kondo scale $T_K^{\alpha}$ (with residual entropy $2\ln(2)$).
For small finite $K<T_K^{R}$, (b), these residual moments form a local singlet 
state on the scale $T\sim K$, so the residual entropy is quenched.
On increasing the interimpurity $K$ further ($T_K^R<K<T_K^L$), the
underscreened spin-$\tfrac{1}{2}$ $L$ impurity and the still
unscreened spin-$1$ $R$ impurity are coupled and form a local doublet
state on the scale $T\sim K$. This can then be single-channel
Kondo screened by an effective coupling either to the $L$ channel (c)
or the $R$ channel (e), and the residual entropy is again
quenched. However, $L$ and $R$ effective couplings can become 
equal on fine-tuning $K$. This is the single spin-$\tfrac{1}{2}$
2CK critical point (d), with residual entropy $\tfrac{1}{2}\ln(2)$. For 
large $K\gg T_K^L$, a local interimpurity singlet state arises as expected (f). 

Analysis of the finite size spectrum at the critical point shows it to
be \emph{identical} to that of the regular spin-$\tfrac{1}{2}$
2IKM, independent of asymmetry~\cite{sm}; and is hence that of a 2CK
model with additional potential scattering.


\emph{Conclusion.--} 
We have shown the critical point of the spin-$S$ 2IKM, including the spin-$\tfrac{1}{2}$ variant,
to be ubiquitously 2CK in nature. However  conductance lineshapes
measurable in experiment exhibit distinctive behavior depending on
underlying symmetries, the low-energy behavior in particular evolving
from square-root to 
linear behavior in $V_{sd}$ or $T$ as the channel symmetric point is approached, and for any spin-$S$.


\emph{Acknowledgments.--} We thank I. Affleck and T. Quella for
insightful input,  
and acknowledge financial support from the DFG through SFB608 and
FOR960 (AKM), the A. v. Humboldt foundation (ES), and EPSRC through
EP/I032487/1 (DEL).


\end{document}


\title{Two-channel Kondo physics in two-impurity Kondo
  models:\\Supplementary Material}

\author{Andrew K. Mitchell,$^{1}$ Eran Sela,$^{1}$ and David E. Logan$^{2}$}

\affiliation{$^{1}$Institute for Theoretical Physics, University of
  Cologne, 50937 Cologne, Germany\\
$^2$Department of Chemistry, Physical and Theoretical
  Chemistry, Oxford University, South Parks Road, Oxford OX1 3QZ,
  United Kingdom}

\maketitle

The following appendices comprise the supplementary material to Ref.~\onlinecite{paper}.

\section{Connecting the finite size spectra of the two-channel and two-impurity Kondo models}

In this appendix we demonstrate that the critical points of the 2CK
model and 2IKM are connected by a marginal operator corresponding to
the potential scattering $H_{ps}$. Specifically, the finite size
spectrum of one model is shown to transform into that of the other
model when this additional term is accounted for explicitly. In the
channel-asymmetric 2IKM, the physical origin of this operator is
readily understood from the mapping~\cite{zarandPRL} between 2IKM and
2CK, where an additional phase shift in the more strongly-coupled
channel arises due to first-stage single-channel Kondo screening of
one of the impurities.

\begin{figure}[h] \begin{center} \includegraphics*[width=85mm]{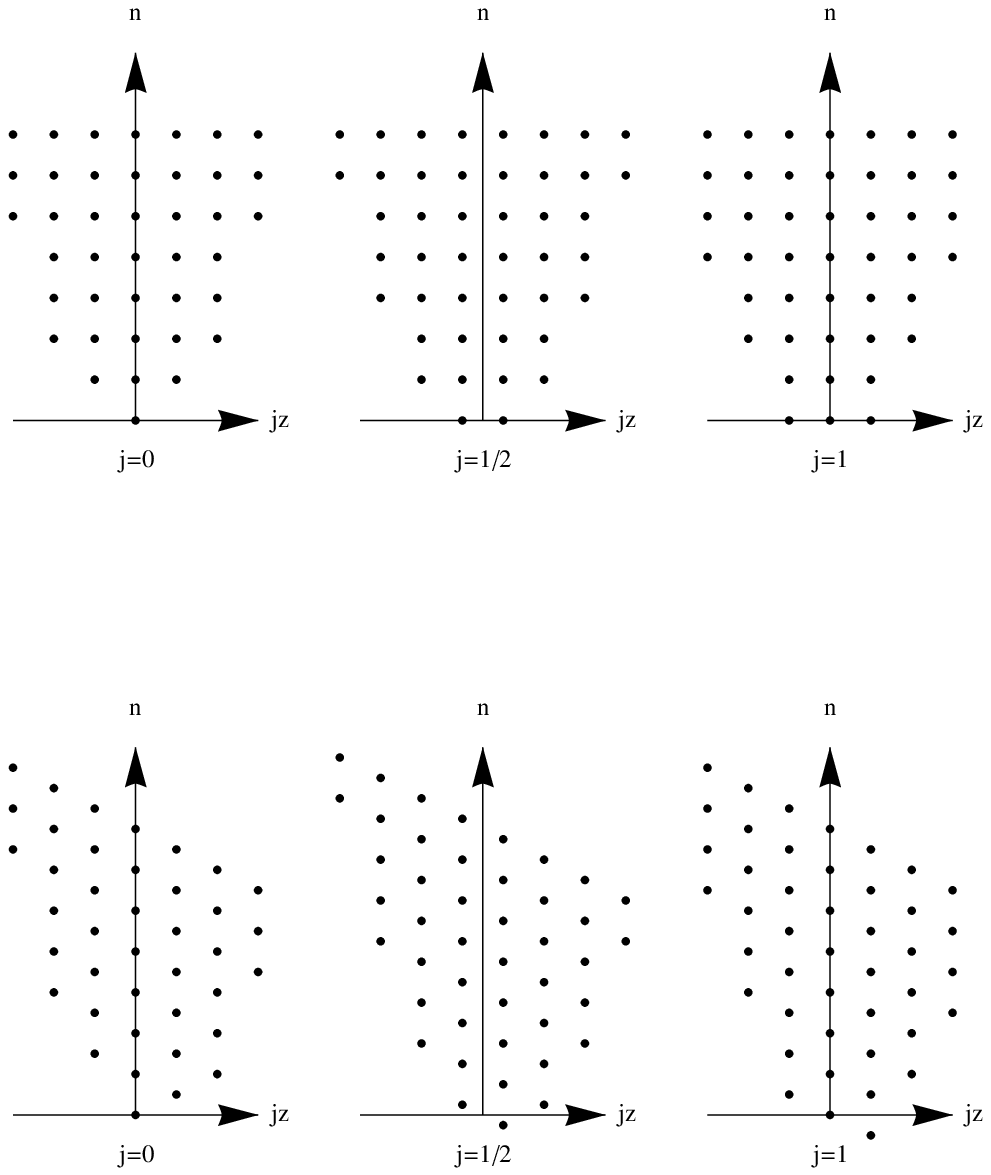}
    \caption{Top panels: $SU(2)_k$ Kac-Moody conformal towers for
      $k=2$, for $j=0,\tfrac{1}{2},1$ from left to right. Bottom panels: tilted towers due to Zeeman term $-b j_F^z$ with $b=\tfrac{1}{2}$. Multiplicities are not shown.}
\label{fg:towers}
\end{center}
\end{figure}

First, we review how the finite size spectrum at the critical point of
the 2CK model is organized according to the unperturbed CFT with 
$U(1) \times SU(2)_2\times SU(2)_2 $ charge, spin and flavor symmetry 
sectors. The spectrum consists of Kac-Moody conformal towers,
obtained by combining these symmetry sectors according to the gluing
conditions\cite{2ck:aff_lud} shown in
Table~\ref{tb:2cknontrivial}. The energy of the 
lowest lying state in each conformal tower (measured in units of
$\frac{2\pi\hbar v_F}{L}$, with $v_F\equiv 1$ the Fermi velocity and
$\hbar\equiv 1$ hereafter) follows from the formula\cite{2ck:aff_lud}
 \be
 \label{spec}
 E_{2CK} = \frac{Q^2}{8}+\frac{j(j+1)}{4}+\frac{j_F(j_F+1)}{4},
 \ee
where $Q$ is the total charge, $j=0,\tfrac{1}{2},1$ is the total spin,
and $j_F=0,\tfrac{1}{2},1$ is the flavor quantum number. Only certain
quantum number combinations are allowed at the 2CK critical
point,\cite{2ck:aff_lud} as given by the nontrivial gluing conditions
of Table~\ref{tb:2cknontrivial} (and with level degeneracies also
following from the Table). Also note that the gluing conditions for
integer $Q$ are defined modulo $2$. 
Each conformal tower has an infinite number of states. Consider for example
the spin $j$ tower of the spin $SU(2)_2$ sector. The states with
lowest energy form a spin $j$ representation of $SU(2)$, and hence are  
$(2j+1)$-degenerate (the corresponding magnetic quantum number is $j_z =
-j,-j+1,...,j$). The reducible representations of $SU(2)$ with
increasing spin (equal to $j$ modulo $1$) give rise to higher energy
states. In the space of $j_z$ and energy $n$, the envelope of an 
$SU(2)_k$ conformal tower on level $k$ with spin $j$ is $
n=[(j^z)^2-j^2]/k$~\cite{gepner}.  For a visualization of such
unperturbed conformal towers, see the upper panels of Fig.~\ref{fg:towers}.

\begin{table}
\begin{center}
\begin{tabular}{|@{$\quad$}r@{$\quad$}r@{$\quad$}r@{$\quad$}r@{$\quad$} r@{$\quad$} |}
\hline
$Q$ & $j$ & $j_F$ & ${\rm{multiplicity}}$ & $E_{2CK}$ \\
\hline
0 & 1/2 & 0 & 2 & $\textstyle{3/16}$  \\
$\pm$ 1 & 0 & 1/2 & 4 & $\textstyle{5/16}$  \\
$\pm$ 1 & 1 & 1/2 & 12 & $\textstyle{13/16}$  \\
$\pm$ 2 & 1/2 & 0 & 4 & $\textstyle{11/16}$  \\
0 & 1/2 & 1 & 6 & $\textstyle{11/16}$  \\
\hline
\end{tabular}
\end{center}
\caption{Nontrivial gluing conditions of conformal towers belonging to the $U(1) \times SU(2)_2
\times SU(2)_2 $ charge, spin and flavor symmetry sectors.} \label{tb:2cknontrivial}
\end{table}

We now perturb the CFT by introducing potential scattering $H_{ps} =
\sum_{\alpha} V_\alpha {\psi^{\dagger}_{0\sigma \alpha}}
\psi^{\phantom{\dagger}}_{0 \sigma \alpha}$. To see how this
perturbation enters in the CFT, we first switch off the Kondo
interactions, and consider free Fermions of a single species 
$\sigma\alpha$. With antiperiodic boundary conditions $\psi_{\sigma
  \alpha}(-L/2) =- \psi_{\sigma \alpha}(L/2)$, the single
particle momenta are $k=\frac{2 \pi}{L} (n+\tfrac{1}{2})$; and for linear
dispersion, the single particle energies are just given by $k$. 
The potential scattering amplitudes $V_{\alpha}$ can be parametrized
in terms of scattering phase shifts $\delta_{\alpha}$. They are
defined in terms of the shift of these single particle levels
$k=\frac{2 \pi}{L} (n+\tfrac{1}{2}-\frac{\delta_{\sigma\alpha}}{\pi})$. For
$N_{\sigma \alpha}$ particles of each species, the correction to the
total energy (again in units of $\tfrac{2\pi}{L}$) is thus 
\be
\label{de}
\Delta E = -\frac{1}{\pi} \sum_{\sigma\alpha}\delta_{\sigma\alpha}N_{\sigma \alpha}.
 \ee
In terms of the particle numbers, the total
charge is $Q = \sum_{\sigma} (N_{\sigma L}+N_{\sigma R})$ while the
magnetic quantum number of the flavor sector is $j_F^z =  \tfrac{1}{2}\sum_{\sigma
 } (N_{\sigma L} - N_{\sigma R})$. In the presence
of $H_{ps}$, the spectrum thus follows as $\tilde{E}_{2CK}=E_{2CK}+\Delta
E$, viz
 \be
 \label{spec1}
 \tilde{E}_{2CK} = \frac{Q^2}{8}+\frac{j(j+1)}{4}+\frac{j_F(j_F+1)}{4}-Q \frac{\delta_L +\delta_R}{2 \pi}-j_F^z \frac{\delta_L -\delta_R}{\pi},
 \ee
where $\delta_{\sigma\alpha}\equiv \delta_{\alpha}$ since the
potential scattering $H_{ps}$ is independent of spin $\sigma$. 
Eq.~(\ref{spec1}) is equivalent to Eq.~(4) of Ref.~\onlinecite{paper}, with
$a=2\frac{\delta_L+\delta_R}{\pi}$ and $b=\frac{\delta_L
  -\delta_R}{\pi}$. Thus $H_{ps}$ simply biases the charge and flavor
conformal towers. For a visualization of the effect on the $SU(2)_2$
flavor towers, see the lower panels of Fig.~\ref{fg:towers}. The
spectrum obtained using the trivial free Fermion gluing
conditions\cite{2ck:aff_lud} describes free electrons but 
with the additional potential scattering. Importantly, switching on
$H_{ps}$ and switching on the Kondo interactions commute, since there
is a spin-charge-flavor separation in the model.\cite{2ck:aff_lud}
Employing instead the nontrivial gluing conditions of
Table~\ref{tb:2cknontrivial} in Eq.~(\ref{spec1}), we thus obtain
the 2CK critical spectrum including the influence of $H_{ps}$.  

\begin{figure}[t] \begin{center} \includegraphics*[width=125mm]{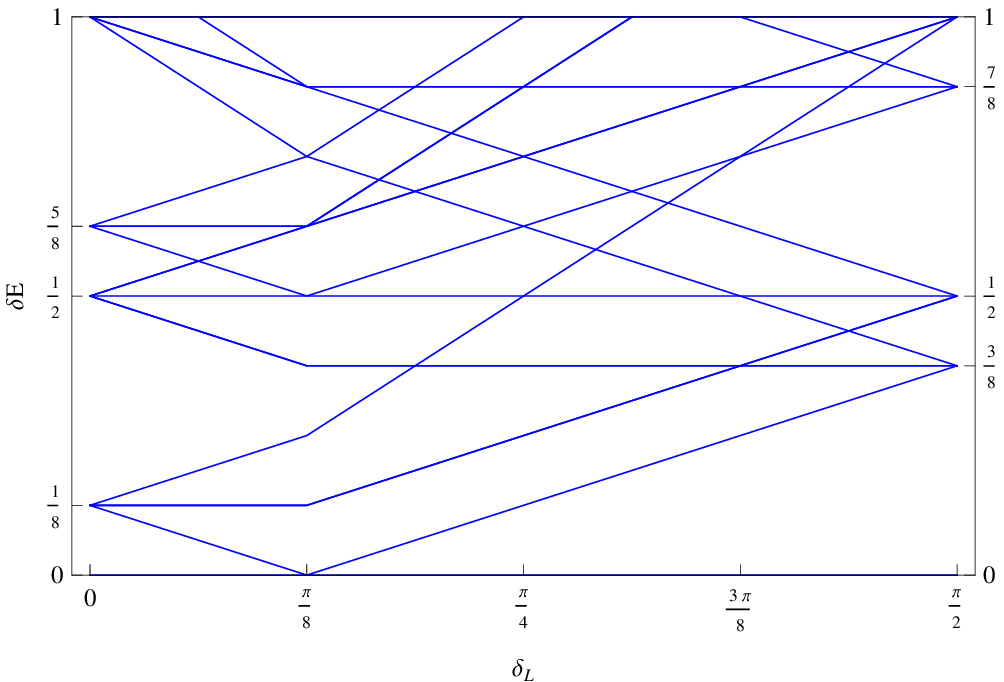}
    \caption{Excitation energies $\delta E$ according to Eq.~(\ref{spec1}) as a 
      function of $\delta_L$ for $\delta_R=0$, connecting the regular
      2CK spectrum ($\delta_L=0$) and the 2IKM spectrum
      ($\delta_L=\pi/2$). Note that at $\delta_L=\pi/8$ the ground
      state changes from the doublet $(Q,j,j_F)=(0,1/2,0)$ to the
      $j_F^z=1/2$, $Q=1$ component of the quintet $(\pm 1,0,1/2)$. }
\label{fg:levels}
\end{center}
\end{figure}

In Fig.~\ref{fg:levels}, we examine the evolution of the energy levels
at the critical point of the 2CK model as function of $\delta_L$ for 
$\delta_R=0$, plotting the lowest excitation energies $\delta 
E_i = \tilde{E}_i - \min \{\tilde{E}_i \}$. The fractions $0$,
$\tfrac{1}{8}$, $\tfrac{1}{2}$, 
$\tfrac{5}{8}$, $1$... correspond to multiplets of the symmetry $U(1)
\times SU(2)_2 \times SU(2)_2 $ of the regular 2CK model. Those
multiplets split due to the flavor field and charge field induced by
finite $H_{ps}$, but eventually at $\delta_L=\pi/2$ they recombine to
give new multiplets characterized by the fractions 
$0$, $\tfrac{3}{8}$, $\tfrac{1}{2}$, $\tfrac{7}{8}$, $1$..., and which
correspond to that of the regular 2IKM.\cite{2ik:aff_lud_jones}

To strengthen the connection between the two models, we now consider
the effect of $H_{ps}$ on the critical CFT of the 2IKM itself. We
exploit here the Bose-Ising representation\cite{2ik:aff_lud_jones} of
the 2IKM, consisting of a decomposition into $SU(2)_1 \times SU(2)_1
\times SU(2)_2 \times Z_2$ left/right charge (isospin), total spin and
$Z_2$ Ising symmetry sectors. The $z$-component of $\alpha=L$ or $R$
isospin is essentially the total charge of channel
$\alpha$. Specifically, its magnetic quantum number is given simply by 
$i_{\alpha}^z=\tfrac{1}{2}\sum_{\sigma}N_{\sigma\alpha}$ in terms of
the particle numbers. Using Eq.~(\ref{de}), it follows that $\Delta E
= -\tfrac{2}{\pi}\sum_{\alpha}\delta_{\alpha}i_{\alpha}^z$. The
energy of the lowest states of each conformal tower in the 2IKM with
the additional potential scattering term $H_{ps}$ is thus
$\tilde{E}_{2IK}=E_{2IK}+\Delta E$, where $E_{2IK}$ is given in
Ref.~\onlinecite{2ik:aff_lud_jones}. Our final result is
\be
 \label{spec2}
 \tilde{E}_{2IK} =  \frac{i_L (i_L+2)}{3} +\frac{i_R (i_R+2)}{3} +\frac{j(j+1)}{4}+ x_{{\rm{Ising}}} - 2 \frac{\delta_L}{\pi} i_L^z-2 \frac{\delta_R}{\pi} i_R^z,
 \ee
where, $i_\alpha=0,\tfrac{1}{2}$ are the isospin quantum numbers, 
$j=0,\tfrac{1}{2},1$ is the total spin quantum number and $x_{Ising} =
0,\tfrac{1}{16},\tfrac{1}{2}$ (corresponding to the unity operator, spin field $\sigma$,
and fermion field $\epsilon$ of the $Z_2$ Ising model). The nontrivial
gluing condition of the $SU(2)_1 \times SU(2)_1  \times SU(2)_2 \times
Z_2$ sectors is given in Table. III of
Ref.~\onlinecite{2ik:aff_lud_jones}, and gives the finite size
spectrum of the 2IKM in the presence of potential scattering.

Using Eqs.~(\ref{spec1}) and (\ref{spec2}) with their corresponding
gluing conditions, we find
\be
\label{rel}
\mathcal{H}_{\textit{CP}}^{2IK}(\delta_L,\delta_R) =
\mathcal{H}_{\textit{CP}}^{2CK}(\delta_L+\pi/2,\delta_R),
 \ee
where $\mathcal{H}_{\textit{CP}}\equiv \{ \delta E_i\}$ at the
critical point of either model, parametrized in terms of the free
Fermion phase shifts $\delta_L$ and $\delta_R$. 
The finite size spectrum at the critical point of one model with
potential scatterings $V_L$ and $V_R$ is identical  
to that of the other model with different $\tilde{V}_L$ and 
$\tilde{V}_R$ when the phase shifts satisfy Eq.~(\ref{rel}). Thus, the 
critical FPs are equivalent in the sense that they lie on the same
marginal manifold parametrized by $H_{ps}$.

\subsection{Alternative formulation in terms of Majorana Fermions}
Having drawn the connection between the non-abelian CFT formulation of
the 2CK and 2IKM, we discuss now the same relation within
the framework of abelian bosonization. 

The steps applied first by Emery and Kivelson\cite{emery} as an
alternative solution to the 2CK model, and later by Gan~\cite{gan} in
the 2IKM context, are as follows. (i) Bosonize separately each chiral
Fermionic species (with the impurities at the `boundary' located at $x=0$), $\psi_{\sigma
  \alpha}(x) \sim e^{-i \phi_{\sigma \alpha}(x)}$. (ii) Define linear
combinations of the bosonic fields $\phi_{\sigma \alpha}(x)$, $\{
\phi_c,\phi_s,\phi_f,\phi_X\}=\frac{1}{2} \sum_{\sigma, \alpha= 1}^2
\phi_{\sigma \alpha}
\{1,(-1)^{\sigma+1},(-1)^{\alpha+1},(-1)^{\sigma+\alpha} \}$, 
corresponding to charge, spin, flavor and spin-flavor bosons
($A=c,s,f,X$). (iii) Define four new Fermionic species by
$\psi_A(x) \sim e^{-i \phi_A(x)}$, whose real
($\chi_1^A=\frac{\psi^\dagger_A + \psi_A}{\sqrt{2}}$) and imaginary
($\chi_2^A=\frac{\psi^\dagger_A - \psi_A}{\sqrt{2}i}$) parts give
eight Majorana Fermions (MFs) $\chi_{j}^A(x)$, (with $j=1,2$ and $A=c,s,f,X$), which may
be regarded as components of the vector $\vec{\chi}(x)$. The
Hamiltonian for the free theory with $J_L=J_R=0$ and 
$H_{\textit{ps}}=0$ is then
$H=\tfrac{i}{2}\int_{-\infty}^{\infty}\textit{dx}~\vec{\chi}(x)\cdot\partial_x\vec{\chi}(x)$, 
with the trivial boundary condition (BC)
$\vec{\chi}(0^+)=\vec{\chi}(0^-)$.  The BC relates the field before 
scattering at $x = 0^+$ and after scattering at $x = 0^-$ (using a
left moving convention). The remarkable fact is that in both 2CK and
2IKM, the critical FP Hamiltonian is the same as that of the free
theory, but with a modified BC that is also simple in terms of the
MFs. For 2CK it is simply\cite{2ck:aff_lud}
\be
\label{bc0}
 \vec{\chi}_s(0^+)=-\vec{\chi}_s(0^-)\qquad \text{and}\qquad \chi_j^A(0^+)= 
 \chi_j^A(0^-)~~\text{for}~~ (A,j) \ne (s,1),(s,2),(X,1),
\ee
with $\vec{\chi}_s = (\chi_1^s,\chi_2^s,\chi_1^X)$, such that the
modified BC affects only the spin sector of the 2CK model. 
For 2IKM,\cite{maldacena}
\be
\label{bc1} 
 \chi_2^X(0^+)=-\chi_2^X(0^-)\qquad \text{and}\qquad \chi_j^A(0^+)= 
 \chi_j^A(0^-)~~\text{for}~~ (A,j) \ne (X,2), 
\ee 
where the modified BC here affects only the Ising sector of the 2IKM.

An \emph{odd} number of MFs suffer the modified BC in both
models. This is a hallmark of NFL behavior, as it
implies\cite{maldacena} a vanishing amplitude for scattering of an
electron into an electron. By contrast, the BCs for Fermi liquids 
always involve an even number of MFs, corresponding thereby to linear
BCs for regular Fermions.  

Besides their common NFL character, the different BCs for
the critical points of the 2CK and 2IKM do not immediately suggest any
connection between the models. However, we now show that adding
potential scattering in one channel of the 2CK model (Eq.~(2) of Ref.~\onlinecite{paper}, with $V_L \to \infty$ and $V_R \to 0$ in the added term $H_{ps}$)
causes a change in the critical 2CK BC which makes it equivalent to
that of the 2IKM.

As above, the addition of infinite potential scattering in the left
channel (which results in the desired $\delta_L=\pi/2$ phase shift) is 
equivalent to the addition of infinite uniform potential scattering 
$\frac{1}{2} \psi_L^\dagger \psi_L +  \frac{1}{2} \psi_R^\dagger
\psi_R = i \chi_2^c \chi_1^c$ which affects only the charge sector,
and infinite staggered potential scattering $\frac{1}{2} \psi_L^\dagger \psi_L -  \frac{1}{2}
\psi_R^\dagger \psi_R = i \chi_2^f \chi_1^f$, affecting only the
flavor sector. Since both charge and flavor sectors are unaffected by the Kondo
interaction (which acts purely in the \emph{spin} sector), the
resulting behavior can be understood from the response of free
Fermions to potential scattering in the simpler situation
$J_L=J_R=0$. For $V_L \to \infty$ and $V_R \to 0$, modified BCs are
conferred to the charge and flavor MFs: 
$\chi_{1,2}^{c,f}(0^+)=-\chi_{1,2}^{c,f}(0^-)$. 
Together with the BC from Eq.~\ref{bc0}, we now obtain the BC for the
critical 2CK model in the presence of the potential scattering, viz
 \be
 \label{bc2}
 \chi_2^X(0^+)=\chi_2^X(0^-)\qquad \text{and}\qquad \chi_j^A(0^+)= -
 \chi_j^A(0^-)~~\text{for}~~ (A,j) \ne (X,2). 
\ee

This BC is the same as that of the 2IKM, Eq.~(\ref{bc1}), except
for an additional minus sign afflicting the BC for all
MFs. However, the BCs corresponding to Eqs.~(\ref{bc1}) and
(\ref{bc2}) are in fact equivalent in the 2CK model, as now shown.

An `overall' minus sign in the BC for the MFs can be interpreted as a
 phase shift felt by each of the charge, spin, flavor and spin-flavor
Fermions, defined in terms of the scattering process
$\psi_A(0^-)=\exp(2i\delta_A)\psi_A(0^+)$, and with 
$\delta_A=\pi/2$ for $A=c,s,f,X$. Since the number operators for
$\psi_A$ Fermions are related\cite{zarandvonDelft} to the number
operators for the regular $\psi_{\sigma\alpha}$ Fermions via $\{
N_c,N_s,N_f,N_X\}=\frac{1}{2} \sum_{\sigma \alpha} N_{\sigma 
  \alpha} \{1,(-1)^{\sigma+1},(-1)^{\alpha+1},(-1)^{\sigma+\alpha}
\}$,  it follows that $\sum_A N_A = 2N_{\uparrow L}$. In terms of the
particle numbers $N_A$, the correction to the total energy due to
additional phase shifts is $\Delta E = -\tfrac{1}{\pi}\sum_A\delta_A
N_A = -N_{\uparrow L}$. Comparing to Eq.~\ref{de}, we then find that
$\delta_{\uparrow L}=\pi$  (and $\delta_{\sigma\alpha}=0$ for
$\sigma\alpha\ne \uparrow L$). Thus the 
additional global minus sign in the BCs for the MFs equates to
$\psi_{\uparrow L}(0^-)=\exp(2i\delta_{\uparrow L})\psi_{\uparrow
  L}(0^+)\equiv \psi_{\uparrow L}(0^+)$ for the original Fermions, and
so has no effect on the finite size spectrum --- or on the
lowest-energy physics in general.

\section{Correction to the Green function near the 2IKM critical 
 point} 
In this appendix we consider corrections to the critical FP
Hamiltonian of the 2IKM, and from them construct corrections to the
Green function. We show that the nature of these corrections (which
show up in physical quantities such as conductance) is characteristic
of the NFL behavior, and depend distinctively on parity breaking
in the model.  

\subsection{Fixed point Hamiltonian of the 2IKM}
First we construct the list of operators allowed by symmetry in the
generic asymmetric 2IKM (see Eq. (1) of Ref.~\onlinecite{paper}), including the
dependence on small $J_L - J_R$ and on small potential scattering
amplitudes $V_L$ and $V_R$. We rely on the same operator
content obtained by Affleck and Ludwig,\cite{2ik:aff_lud_jones} based
on the $SU(2)_1 \times SU(2)_1  \times SU(2)_2 \times Z_2$ symmetry
decomposition. The critical FP of the symmetric 2IKM 
without potential scattering has only one irrelevant
operator of dimension $3/2$, denoted $\epsilon'$, and is given by Eq.~(5.1)
of Ref.~\onlinecite{2ik:aff_lud_jones}. In the present context where parity 
and particle-hole symmetries are in general broken, additional operators 
are allowed, and in fact dominate the approach to the NFL FP.  

Each operator that can occur at an unstable FP is either a
primary operator or a descendant of it.\cite{2ik:aff_lud_jones}
The primary operators are labeled with the same quantum numbers 
$(i_1,i_2,j,{\rm{Ising}})$ used to label states in
Eq.~(\ref{spec2}). The particular combinations of $SU(2)_1 \times
SU(2)_1  \times SU(2)_2 \times Z_2$ left/right isospin, total spin and
Ising operators, together with their scaling dimension $x$, was obtained  
using the double fusion method in Ref.~\onlinecite{2ik:aff_lud_jones}. For 
clarity and completeness we reproduce that list here in Table \ref{tb}.  

\begin{table}
\begin{center}
\begin{tabular}{|@{$\quad$}r@{$\quad$}r@{$\quad$}r@{$\quad$}r@{$\quad$} r@{$\quad$} |}
\hline
$i_L$ & $i_R$ & $j$ & ${\rm{Ising}}$ & $x$ \\
\hline
$0$  & $0$ &  $0$ & $1$ & $0$  \\
$\textstyle{\frac{1}{2}}$ & $0$ & $\textstyle{\frac{1}{2}}$ & $\sigma$ &  $\textstyle{\frac{1}{2}}$ \\
$0$ & $\textstyle{\frac{1}{2}}$ & $\textstyle{\frac{1}{2}}$ & $\sigma$ &  $\textstyle{\frac{1}{2}}$ \\
$0$  & $0$ & $1$ & $1$ & $\textstyle{\frac{1}{2}}$ \\
$0$  & $0$ & $0$ & $\epsilon$ & $\textstyle{\frac{1}{2}}$  \\
$\textstyle{\frac{1}{2}}$  & $\textstyle{\frac{1}{2}}$ &  $0$ & $1$ & $\textstyle{\frac{1}{2}}$  \\
$0$  & $0$ & $1$ & $\epsilon$ & $1$  \\
$\textstyle{\frac{1}{2}}$  & $\textstyle{\frac{1}{2}}$ & $1$ & $1$ & $1$ \\
$\textstyle{\frac{1}{2}}$  & $\textstyle{\frac{1}{2}}$ & $0$ & $\epsilon$ & $1$ \\
$\textstyle{\frac{1}{2}}$  & $\textstyle{\frac{1}{2}}$ & $1$ & $\epsilon$ & $\textstyle{\frac{3}{2}}$\\
\hline
\end{tabular}
\end{center}
\caption{Operator content of the 2IKM critical FP. } \label{tb}
\end{table}

In general, the presence of a given symmetry implies that the only
allowed operators are those which are singlets of that symmetry.  In
particular, the full $SU(2)_1\times SU(2)_1  \times SU(2)_2 \times
Z_2$ symmetry of the standard 2IKM with $H_{ps}=0$ gives rise to the
leading singlet operator $ (i_1,i_2,j,{\rm{Ising}}) =
(0,0,0,\epsilon)$. This relevant operator is
associated\cite{2ik:aff_lud_jones} with the perturbation $(K-K_c)$,
which destabilizes the 2IKM critical point. At the critical point
$K=K_c$, one leading irrelevant operator\cite{2ik:aff_lud_jones} is
the descendant of $(0,0,0,\epsilon)$, denoted 
$\mathcal{O}_1 =\epsilon' \approx \partial_\tau \epsilon$. Since
$\epsilon$ has scaling dimension $1/2$, its first descendant has
scaling dimension $3/2$. But singlets can also be obtained as
descendants of primary $SU(2)$ fields with integer spin. Indeed, the
other singlet operator with dimension $3/2$ is in this case 
obtained by acting on the $\vec{\phi}=(0,0,1,1)$ primary vector
field of the $j=1$ spin sector with an operator that creates spin
excitations.\cite{2ik:aff_lud_jones} Thus, one must consider also a
second irrelevant operator $\mathcal{O}_2 = \vec{J}_{-1} \cdot \vec{\phi}$
(where $\vec{J}_{-1}$ is the lowest Fourier mode of the spin current
creating such excitations).

Since $\vec{J}_{-1} \cdot \vec{\phi}$ is also the leading irrelevant
operator in the 2CK model,\cite{CFT2CK} one might expect the physical
behavior stemming from it to be common to both 2CK and 2IKM. However,
the crucial difference between the models is that $\vec{J}_{-1} \cdot
\vec{\phi}$ has odd parity in the 2IKM but even parity in the 2CK
model; as now shown.

Parity symmetry (corresponding to left/right reflection in space) is
defined by invariance under the permutations $\vec{S}_L \leftrightarrow
\vec{S}_R$ and $ \psi_L \leftrightarrow \psi_R$. One rather subtle
consequence\cite{2ik:aff_lud_jones} of parity symmetry in the 2IKM is
that $\vec{\phi} \to - \vec{\phi}$. This can be seen from the
parity-odd operator $\psi_L^{\dagger} \vec{\sigma} \psi_L^{\phantom{\dagger}} - \psi_R^{\dagger} \vec{\sigma} \psi_R^{\phantom{\dagger}} = \vec{\phi}
\epsilon$ (and noting that $\epsilon$ is the operator corresponding to
$(K-K_c)$, and is as such unaffected by the reflection,
$\epsilon\leftrightarrow \epsilon$). Thus $\vec{J}_{-1} \cdot
\vec{\phi}$ is also parity-odd in 
the 2IKM. But in the 2CK model, $\psi_L^{\dagger} \vec{\sigma} \psi_L^{\phantom{\dagger}} - \psi_R^{\dagger} \vec{\sigma} \psi_R^{\phantom{\dagger}} =
\vec{\phi} \phi_F^z$. Here, $\vec{\phi}_F$ is the $j_F=1$ primary
field of the flavor sector, and $\phi_F^z$ is its $z$-component. Since
parity corresponds to a $\pi$ rotation around the $x$ flavor direction
in the 2CK model, the flavor current $\vec{J}_F =
\sum_{\sigma} \psi_{\sigma}^{\dagger} \frac{\vec{\tau}}{2}
\psi_{\sigma}^{\phantom{\dagger}}$ transforms as $J_F^x \to J_F^x$,
$J_F^{y,z} \to - J_F^{y,z}$; and similarly $\phi_F^x \to \phi_F^x$,
$\phi_F^{y,z} \to -\phi_F^{y,z}$. Thus $\vec{J}_{-1} \cdot \vec{\phi}$
is parity even in the 2CK model. Of course, this is the expected
result here since $\vec{J}_{-1}$ and $\vec{\phi}$ are both spin
operators, but the parity transformation affects only the flavor
sector of the 2CK model.

Finally, we consider the additional operators appearing when isospin
$SU(2)_1 \times SU(2)_1 $ symmetry is broken (but spin $SU(2)$
symmetry is maintained), as occurs when potential scattering from
$H_{ps}$ is included. Specifically, $H_{ps} =2 \sum_{\alpha} V_\alpha
I_\alpha^z(x=0) $, where the $z$-component of the channel-$\alpha$
isospin current is given by
$I_{\alpha}^z(x)=\tfrac{1}{2}\sum_{\sigma}\psi_{\sigma\alpha}^{\dagger}(x)\psi_{\sigma\alpha}^{\phantom{\dagger}}(x)$. Since
$V_{\alpha}$ does not couple to fields with half-integer isospin, and
the primary operators have total spin $j=0$, then it follows that
$V_{\alpha}$ couples only to dimension $3/2$ descendants of the
$(0,0,0,\epsilon)$ field. Thus, two further dimension $3/2$ operators
$\mathcal{O}_{3,4} = (I^z_{L,R})_{-1} \epsilon$ are obtained (where 
$(\vec{I}_{\alpha})_{-1}$ is the isospin analogue of the spin operator
$\vec{J}_{-1}$ discussed above). No other dimension $3/2$ operators
are consistent with the symmetry. 

The corrections to the critical FP Hamiltonian in the presence of
$H_{ps}$ thus comprise four operators,
\be
\label{deltaS}
\delta H = \frac{1}{\sqrt{T_c}}\sum_{i=1}^4 c_i \mathcal{O}_i.
\ee
As shown in Ref.~\onlinecite{2ik:aff_lud_jones}, the coefficient $c_1
\sim 1$. By contrast, the odd transformation property of
$\mathcal{O}_2$ under parity implies 
\begin{equation}
\label{c2var}
c_2 \sim \left (\frac{J_L - J_R}{J_L +
  J_R}\right )
\end{equation}
for small small $(J_L - J_R)$. Finally $c_{3,4}
\propto V_{L,R}$ for small $V_{L,R}$. 

In the conformal limit (which relies upon linear dispersion and infinite conduction bandwidth), the Bose-Ising representation of the 2IKM reads $H_{2IK} = H_0+\vec{J}(0) \cdot (J_L \vec{S}_L+J_R
\vec{S}_R)+\vec{\phi}(0)\epsilon \cdot (J_L \vec{S}_L-J_R \vec{S}_R)+ 2
\sum_{\alpha} V_\alpha I_\alpha^z(0) $, showing that there is perfect separation of isospin and Ising sectors (although there is a coupling between Ising and spin via the term $\vec{\phi}(0)\epsilon$). Since $\mathcal{O}_{3,4}$ involve coupling between isospin and Ising operators $(I^z_{L,R})_{-1}$ and $\epsilon$, in this idealized limit one thus obtains $c_{3,4} =0$ identically.   
But in generic models, one naturally expects $c_{3,4} \ll 1$ to be finite but small. For this reason we ignore the operators $\mathcal{O}_{3,4}$ in the present work.

However, we do note that the $\mathcal{O}_{3,4}$  operators do play a significant role in variants of the standard 2IKM which contain explicit coupling between isospin and Ising or spin sectors. The two-impurity Anderson model is a pertinent example: the low-energy effective model\cite{2ik:aff_lud_jones,DEL2iam} is a 2IKM but with additional terms such as $\psi^\dagger (\vec{\sigma} \tau^x) \psi\cdot (\vec{S}_L+\vec{S}_R)$. These terms mix the spin and isospin sectors, leading thereby to indirect coupling between $(I^z_{L,R})_{-1}$ and $\epsilon$ operators, and hence in such situations $c_{3,4}$ are not expected to be small.

\subsection{Corrections to the Green function}
Corrections to the t matrix and conductance in the vicinity of the 2IKM critical point are directly related to corrections to the single particle Green function.\cite{CFT2CK}  
Ignoring the operators $\mathcal{O}_{3,4}$ as above, the leading correction to the Green function to first order in Eq.~(\ref{deltaS}) is $\delta G(z_1,z_2)=\delta_{1} G(z_1,z_2)+\delta_{2} G(z_1,z_2)$, where
\bea
\delta_1 G(z_1,z_2) &=& \frac{c_1}{\sqrt{T_c}} \int_0^\beta d \tau \langle \psi(z_1)  \epsilon'(0,\tau) \psi^\dagger (z_2) \rangle, \nonumber \\
\delta_2 G(z_1,z_2) &=& \frac{c_2}{\sqrt{T_c}} \int_0^\beta d \tau \langle \psi(z_1) \vec{J}_{-1} \cdot \vec{\phi}(0,\tau) \psi^\dagger (z_2) \rangle,
\eea
and $\beta = 1/T$ is inverse temperature. Here we suppress the spin and channel indices, and use $z_1 = \tau_1+i r_1$, $\bar{z}_2 = \tau_2 + i r_2$. Assuming $r_1>0$ and $r_2<0$, the propagator is sensitive to scattering from the impurities at the boundary located at $r=0$. The calculation of $\delta_2 G(z_1 ,z_2)$ was performed by Affleck and Ludwig,\cite{CFT2CK} who exploited the fact that the three-point function in $\delta_2 G(z_1,z_2)$ is fully determined by conformal invariance (up to an overall constant) since $\vec{J}_{-1} \cdot \vec{\phi}$ is a Virasoro primary. The electron and Ising fields, $\psi$ and $\epsilon$, are chiral fields with scaling dimension $1/2$, and thus\cite{CFT2CK}
\bea
\label{tpf}
\delta_2 G(z_1 ,z_2) \propto   \frac{c_2}{\sqrt{T_c}}   \left( \frac{\pi}{\beta}\right)^{5/2}  \times  \nonumber \\
    \int_0^\beta d \tau \frac{[\sin \frac{\pi}{\beta} (z_1 -z_2)]^{1/2}}{\left[\sin \left( \frac{\pi}{\beta} (\tau -z_1) \right) \sin \left( \frac{\pi}{\beta} (\tau -z_2) \right) \right]^{3/2}}.
\eea
The $\tau$ integral can be expressed in terms of hypergeometric functions, which yield asymptotically the famous $\sqrt{\omega}$ energy-dependence of the related t matrix.\cite{CFT2CK}

The integrand in $\delta_1 G(z_1 ,z_2)$ is similarly a three point function, but $\epsilon' \approx \partial_\tau \epsilon$ is not a Virasoro primary field. However, $\epsilon$ itself is Virasoro primary. Pulling the derivative out of the correlation function, one obtains
\bea
\label{regular}
\delta_1 G(z_1,z_2) =\frac{c_1}{\sqrt{T_c}} \int_0^\beta d \tau \partial_\tau \langle \psi(z_1)  \epsilon(0,\tau) \psi^\dagger (z_2) \rangle \propto \nonumber \\
\frac{c_1}{\sqrt{T_c}}  [\sin \frac{\pi}{\beta} (z_1 -z_2)]^{-1/2} \left(\frac{\pi}{\beta} \right)^{3/2}\int_0^\beta d \tau \partial_\tau \frac{1}{\left[\sin \left( \frac{\pi}{\beta} (\tau -z_1) \right) \sin \left( \frac{\pi}{\beta} (\tau -z_2) \right) \right]^{1/2}}=0.
\eea
 The integrand has no singularities as long as $z_1$ and $z_2$ are away from the boundary, and no branch cuts associated with the square-root function are intersected. Periodicity in $\beta$ then implies that the integral vanishes, and so the contribution to
 the Green function from $\delta_1 G(z_1,z_2)$ also vanishes. Thus, the anomalous square-root correction to the Green function must come from $\delta_2 G(z_1,z_2)$ alone. In particular, our conclusion is that the coefficient of the square-root energy-dependence of the t matrix and hence conductance is proportional to $c_2$, which vanishes in the symmetric 2IKM, obtained as $J_L \to J_R$.

Higher-order corrections to the Green function can be calculated in a
similar fashion. In particular, the correction coming from the next
order in perturbation theory involves integrals such as $ \int_0^\beta
d \tau \int_0^\beta d \tau'  \langle \psi(z_1)
\mathcal{O}_i(0,\tau)\mathcal{O}_i(0,\tau') \psi^\dagger (z_2)
\rangle$. Since such integrals contains singularities when $\tau \to
\tau'$, the correction does not in general vanish, even when
$\partial_{\tau}$ and $\partial_{\tau'}$ are pulled outside the
correlator in the case of $\mathcal{O}_1=\epsilon'$. Such calculations
are notoriously involved; and here necessitate the use of an
ultraviolet cutoff $\mathcal{O}(T_c)$ to avoid unphysical divergences. 
The leading correction is however expected to yield a \emph{linear} 
energy dependence (albeit up to possible $\log$ corrections), in
agreement with our NRG calculations, see Fig.~2 of Ref.~\onlinecite{paper}.


\section{2IKM critical physics within the numerical renormalization group}
\label{nrg}

\begin{figure}[t] \begin{center} \includegraphics*[width=160mm]{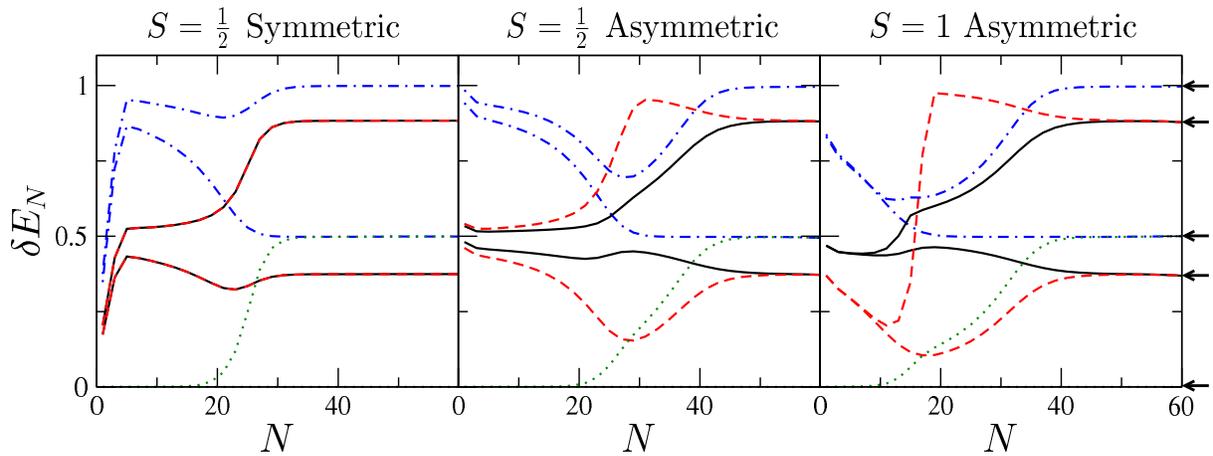}
    \caption{
NRG energy levels $\delta E_N$ for odd iteration number, N. The lowest
2 levels are shown in each left/right charge and total spin subspace
with quantum numbers $(Q_L,Q_R,S^z_{\text{tot}})=(0,0,0)$ [dotted
lines], $(0,1,\tfrac{1}{2})$ [solid lines], $(1,0,\tfrac{1}{2})$
[dashed lines] and $(1,1,0)$ [dot-dashed lines]. \emph{Left panel}:
symmetric spin-$\tfrac{1}{2}$ 2IKM with $\rho J_L=\rho
J_R=0.075$. \emph{Middle panel}: asymmetric spin-$\tfrac{1}{2}$ 2IKM
with $\rho J_L=0.075$ and $\rho J_R=0.05$. \emph{Right panel}:
asymmetric spin-$1$ 2IKM with $\rho J_L=0.15$ and $\rho
J_R=0.05$. Interimpurity coupling $K\simeq K_c$ tuned to its critical value
in each case, and $H_{ps}=0$.
}
\label{fig:nrglevels}
\end{center}
\end{figure}

In this appendix we consider the role of parity breaking on the RG
flow, finite size spectrum and thermodynamics of the 2IKM. Such 
information can be extracted from NRG calculations,\cite{wilson_nrg}
as now briefly reviewed.

Since its first application to the single-impurity single-channel
Kondo model,\cite{wilson_nrg} Wilson's NRG technique has been used
successfully to study a wide range of quantum impurity problems (for a
comprehensive recent review, see Ref.~\onlinecite{NRG_rev}). More
recently, the increase in computing resources has permitted the
detailed analysis of two-channel models involving several impurities,
such as the 2IKM considered in the present work. 

The key element of the NRG technique\cite{wilson_nrg,NRG_rev} is a logarithmic
discretization of the free conduction electron Hamiltonian, $H_0$. The
continuum of states in each conduction band is divided into intervals
with discretization points $x_n=\pm D\Lambda^{-n}$ (here 
$n=0,1,2,...$ and $2D$ is the bandwidth), and whose width thus decreases 
exponentially as the Fermi level is approached. A single state (the
symmetric linear combination) is then retained in each interval, such that
low-energies are exponentially-well resolved. Canonical transformation
by tridiagonalization yields the Wilson chain
representation,\cite{wilson_nrg,NRG_rev} 
where each conduction channel corresponds to a semi-infinite chain
terminated by the impurity sub-system. The discretized Hamiltonian is
then diagonalized iteratively: 
starting from the impurity, Wilson chain orbitals are coupled on
successively and the system diagonalized. To avoid exponential 
growth of the Hilbert space, high-energy states are discarded after
each step. This truncation scheme correctly allows calculation of the
lowest-energy eigenenergies $\{ E_N \}$ of a finite Wilson chain of length $N$
because the coupling between the Wilson chain orbitals $N$ and $(N+1)$
scale as $\Lambda^{-N/2}$. In consequence,\cite{wilson_nrg,NRG_rev}
high-lying states at one 
iteration do not cross over and become low-lying states at a later
iteration due to the energy scale separation inherent when $\Lambda>1$.

$\delta E_N(i)  = \Lambda^{N/2}(E_N(i)-\min\{E_N\})$ are rescaled
many-particle energies (indexed $i$ and measured with respect to the 
ground state energy), and span roughly the same energy range,
independent of $N$. The evolution of these levels with $N$ can be
understood in terms of an RG flow,\cite{wilson_nrg,NRG_rev} with the
various fixed points 
giving characteristic spectra which do not change upon further
iteration. Indeed, the lowest energy levels at
the stable fixed point (obtained as $N\rightarrow \infty$) reproduce
accurately the finite size spectrum obtained by CFT.\cite{note:fssnrg}

Thermodynamics can also be calculated from these NRG energy
levels.\cite{wilson_nrg,NRG_rev}  The essential step here is the
identification of a 
characteristic temperature $T_N\sim \Lambda^{-N/2}$, at which
thermodynamic quantities can be accurately calculated for a given
finite iteration $N$. This temperature is chosen to be high enough
that the splitting of the levels incurred at later iterations does not
affect the thermodynamic calculation; but not too high that states
above the truncation limit contribute significantly.  Thus useful
physical information can be extracted from each iteration, and so the
full temperature-dependence of thermodynamic quantities can be built
up.

\subsection{Effect of parity-braking on RG flow and finite size spectrum}

In Fig.~\ref{fig:nrglevels} we show the evolution of low-lying NRG
eigenenergies $\delta E_N$ with iteration number (Wilson chain length)
$N$, for 2IKMs at criticality. 
Specifically, we compare the channel-symmetric and -asymmetric
spin-$\tfrac{1}{2}$ 2IKM and the asymmetric spin-$1$ 2IKM (with
$H_{ps}=0$ in each case). The RG flow, as evidenced by the flow of 
these levels, is manifestly different for the three cases
considered. In particular, levels with left/right charge and total
spin quantum numbers $(Q_L,Q_R,S^z_{\text{tot}})=(0,1,\tfrac{1}{2})$ and
$(1,0,\tfrac{1}{2})$ are of course degenerate in the symmetric 2IKM
(left panel), since the Hamiltonian is invariant on swapping $L$ and
$R$ labels. By contrast no such symmetry of the bare Hamiltonian is
present in the channel-asymmetric spin-$\tfrac{1}{2}$ or spin-$1$
2IKMs plotted in the middle and right panels; and thus levels related
by $L\leftrightarrow R$ permutation are not in general degenerate.

Importantly however, an emergent parity symmetry at the stable NFL
fixed point is observed, with $(Q_L,Q_R,S^z_{\text{tot}})$ and
$(Q_R,Q_L,S^z_{\text{tot}})$ levels becoming degenerate as
$N\rightarrow \infty$.

Indeed, the set of NFL fixed point levels in each case is identical,
demonstrating that the stable fixed point itself is 
identical, irrespective of bare model symmetries, and independent of
spin-$S$. The fixed point levels are indicated by the arrows, and
correspond to the fractions $0$, $\tfrac{3}{8}$, 
$\tfrac{1}{2}$, $\tfrac{7}{8}$, $1$, $...$, as obtained for the regular
symmetric 2IKM by CFT.\cite{2ik:aff_lud_jones}

\begin{figure}[t] \begin{center} \includegraphics*[width=100mm]{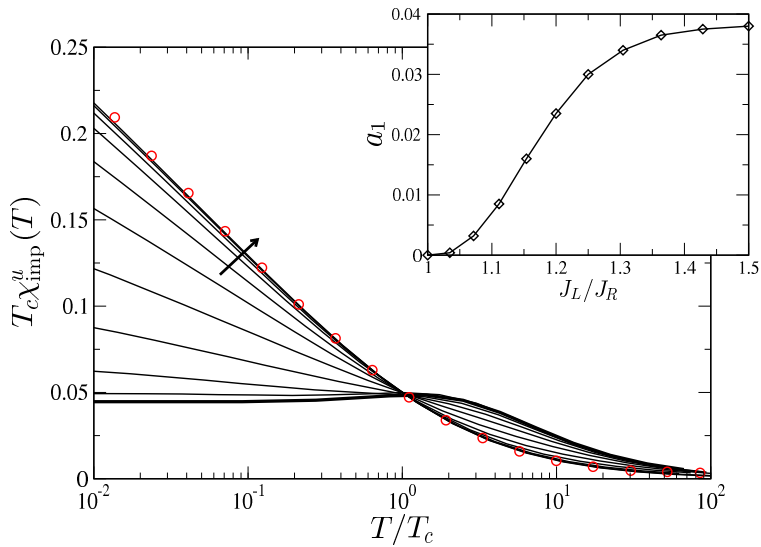}
    \caption{
Uniform magnetic susceptibility $T_c\chi_{\text{imp}}^u(T)$ vs
temperature $T/T_c$ for the 2IKM model. Shown for  $\rho J_L=0.075 \ge
\rho J_R$, varying $\rho 
J_R=0.075\rightarrow 0.05$ in steps of $0.0025$, with $K=K_c$ retuned
in each case (and $H_{ps}=0$), as in Fig. 2 of Ref.~\onlinecite{paper}. Thick solid
line is the symmetric case 
$J_L=J_R$; asymmetry $J_L/J_R\ge 1$ increases in direction of the
arrow. Circle points for a 2CK model with $T_K^{2CK}\equiv T_c$. Inset
shows the variation of the slope $a_1$ as a function of $J_L/J_R$
(see Eq.~\ref{chilog}). $a_2\approx 0.05$ is essentially independent
of asymmetry. We have defined $T_c$ here through
$T_c\chi_{\text{imp}}^u(T_c)=0.05$.
}
\label{fig:nrgchi}
\end{center}
\end{figure}

\subsection{Effect of parity-braking on thermodynamics}

We turn now to thermodynamics, focusing on the `impurity'
contribution\cite{wilson_nrg,NRG_rev} to the uniform
spin susceptibility, $\chi_{\text{imp}}^u(T) = \langle
  (\hat{S}^z_{\text{tot}})^2\rangle_{\text{imp}}/T$ (here
$\hat{S}^z_{\text{tot}}$ refers to the spin of the entire system and 
$\langle \hat{\Omega}\rangle_{\text{imp}} = \langle
\hat{\Omega}\rangle - \langle
\hat{\Omega}\rangle_0$, with $\langle
\hat{\Omega}\rangle_0$ denoting a thermal average in the
absence of the impurities).

In the 2CK model, the uniform susceptibility diverges
logarithmically\cite{andrei,trsvelik} 
at low temperatures $T\ll T_K^{2CK}$,
\begin{equation}
\label{chilog}
T_K^{2CK}\chi_{\text{imp}}^u(T)=a_1\ln(T_K^{2CK}/T) +a_2.
\end{equation}
One naturally expects the uniform susceptibility of the 2IKM to behave
similarly in the channel-asymmetric limit (with $T_K^{2CK}\equiv
T_c$), since here there is a 
mapping to the 2CK model.\cite{zarandPRL} However, as pointed out in
Ref.~\onlinecite{2ik:aff_lud_jones}, the uniform susceptibility is \emph{not}
singular in the regular symmetric 2IKM. Rather, it is the staggered
susceptibility that is divergent in this case.

In Fig.~\ref{fig:nrgchi} we show how the uniform susceptibility
$T_c\chi_{\text{imp}}^u(T)$ vs temperature $T/T_c$ evolves with
increasing asymmetry for the spin-$\tfrac{1}{2}$ 2IKM. In the
case of large channel asymmetry ($J_L/J_R=1.5$), the behavior is
indeed that of Eq.~(\ref{chilog}), with coefficients $a_1$ and $a_2$
essentially those of the regular 2CK model (see comparison to the pure
2CK case, circle points). But, in analogy to
the vanishing square-root energy dependence of conductance at the
symmetric point, we find that the coefficient $a_1\rightarrow 0$ as
$J_L/J_R\rightarrow 1$ (see inset). Indeed, the leading contribution
to the uniform susceptibility in this limit can be
understood\cite{2ck:aff_lud} from second-order perturbation theory in
the leading irrelevant operator 
$\mathcal{O}_2=\vec{J}_{-1} \cdot \vec{\phi}$. From Eq.~(\ref{c2var})
it then follows that $a_1\propto (J_L-J_R)^2$ for small $(J_L-J_R)$. 
Thus, there is a smooth crossover between the limiting cases, with
divergent 2CK behavior arising in the 
asymmetric limit, but constant uniform susceptibility as
$T\rightarrow 0$ emerging at the symmetric point, consistent 
with Ref.~\onlinecite{2ik:aff_lud_jones}.